# Responding to student feedback: Individualising teamwork scores based on peer assessment

**Homa Babai Shishavan [1] and Mahdi Jalili [1] and**

1 Monash University, Melbourne, Australia

2 School of Engineering, RMIT University, Melbourne, Australia

Email for correspondence: homa.babai@monash.edu, mahdi.jalili@rmit.edu.au

**Abstract**

Teamwork assessments often result in a single final product, for which all team members receive a single mark regardless of their contribution to the team project. In order to respond to feedback from students in terms of fair marking of the teamwork assessment, we implemented peer assessment as a recommended pedagogical intervention to individualise team marks and prevent team members from taking advantage of free riding. Team marks were individualised proportional to the average peer assessment mark each student received from their peers in the team. We analysed course evaluation data before ($n = 151$) and after ($n = 129$) the implementation of the peer assessment process from students participating in Engineering Design courses. Analysing data in light of Social Interdependence Theory (Johnson & Johnson, 2009) showed that the peer assessment process improved the cooperation of the team members which enhanced the students' teamwork experience and their engagement with the course. While only 24% of the students appreciated teamwork aspects of the course before the peer assessment, it increased to 34% post peer assessment. Furthermore, the implementation of the peer assessment process decreased complains about free riding from 26% to 7%.

## Introduction

Acquiring efficient teamwork skills is an important part of courses in higher education as a desirable graduate attribute and an important employability skill. Both educators and employers encourage teamwork as an important skill required to undertake a professional job (Budimac, Putnik, Ivanović, Bothe, & Schuetzler, 2011; Iacob & Faily, 2019; Riebe, Girardi,

& Whitsed, 2016). Developing teamwork skill is particularly important in engineering education, where students should acquire the skills to work as a part of a team to improve their engineering problem solving skills (Bae, Ok, & Noh, 2019; Lingard & Barkataki, 2011; Weinberg, White, Karacal, Engel, & Hu, 2005).

Engineering courses in Australian higher education institutes are accredited by Engineers Australia, a member of Washington Accord, a non-for-profit and professional national organisation dedicated to establishing a forum for advancement and standardisation of engineering disciplines within Australia. Engineers Australia frequently review engineering degrees offered by Australian universities and accredit the eligible ones. To be accredible, the universities must show that the courses offered in their engineering programs provide the skillset outlined in "Stage 1 competency for professional engineer" (Engineers Australia, 2019). As part of the elements and indicators of engineering application ability, graduates are required to "Contribute(s) to and/or manage(s) complex engineering project activity, as a member and/or as the leader of an engineering team." One of the six professional and personal attributes is "Effective team membership and team leadership", where graduates are required to:

a) understand the fundamentals of team dynamics and leadership;

b) function as an effective member or leader of diverse engineering teams, including those with multi-level, multi-disciplinary and multi-cultural dimensions;

c) earn the trust and confidence of colleagues through competent and timely completion of tasks;

d) recognise the value of alternative and diverse viewpoints, scholarly advice and the importance of professional networking;

e) confidently pursues and discerns expert assistance and professional advice; and,

f) take initiative and fulfil the leadership role whilst respecting the agreed roles of others.

Including the above skillset in the accreditation process, emphasises of the importance of integrating cooperative learning into the curriculum to develop the necessary teamwork skills for engineering graduates.

While courses with a focus on building teamwork skills are critical in engineering programs, assessing teamwork projects can be challenging. One of the common challenges in assessing teamwork projects is that it usually is not easy to account for the team members individual contributions to the accomplishment of the team project (Friess & Goupee, 2020) often due to the lack of objective evidence (Fidalgo-Blanco, Sein-Echaluce, García-Peñalvo,

& Conde, 2015). Giving a single mark for the team product can be the easiest way of solving this problem. However, the drawback of this approach is that everyone in the group receives the same score regardless of their contribution to the project. In other words, it does not consider the problem of 'free-riding' or 'social loafing' which has been identified as one of the most important negative student behaviours in teamwork (Borrego, Karlin, McNair, & Beddoes, 2013). Research has also shown that students highly favour receiving individualised assessment marks over all group members receiving a single mark for the team work (Tucker, Fermelis, & Palmer, 2009). Therefore, it is necessary to individualize the group marks in an efficient and fair manner that is fairly simple and easy for the students to understand (Gransberg, 2010). This is especially of high importance in engineering and computer science subjects where the outcome of teamwork is usually a single product.

Peer evaluation has been recommended as a reliable pedagogical intervention to minimise social loafing (Borrego et al., 2013; Gransberg, 2010; Gueldenzoph & May, 2002; Mellor, 2012). Peer evaluation, refers to students formally assessing the performance of their team members in the team activity (Topping, 2009). As a student centred assessment practice, it enables the assessors to implement students' input in the assessment process (Pond & Ul-Haq, 1997). This is important because the activities leading to the completion of the teamwork usually are carried out outside the classroom in the absence of the assessing academics and the students involved in the teamwork have a more in-depth insight into the team processes. Therefore, the students' assessment of the team dynamics can reflect a more accurate account of the individual team member's contributions to the completion of the teamwork (Tucker et al., 2009).

Peer assessment is also recommended as one of the strategies to enhance cooperative learning (Strom & Strom, 2011). Successful cooperative learning requires productive cooperation of the group members. Based on social interdependence theory, effective cooperation is facilitated by five variables, which are positive interdependence, individual and group accountability, promotive interaction, social and interpersonal skills, and group processing (Johnson & Johnson, 2009). Positive interdependence exists when the goal achievement of individuals are positively correlated and their actions are driven by the perception that they will achieve their goal only if their group members are able to fulfil their goals (Johnson, 2003; Johnson & Johnson, 2009). Positive interdependence consequentially creates sense of group and individual accountability and responsibility, which motivate group members to perform well by not only trying to complete their own share of work, but also mediating the completion of the work of other group members (Johnson, 2003; Johnson &

Johnson, 2009). Furthermore, positive interdependence leads to promotive interaction among group members and encourages them to provide each other with necessary assistance, share and exchange relevant materials and resources, and improve each other's work by providing constructive feedback (Johnson, 2003; Johnson & Johnson, 2009). Effective promotive interaction among individuals in turn requires their appropriate use of social and interpersonal skills(Johnson & Johnson, 2009). Finally, group processing is an essential element of effective cooperation among individuals where they reflect on the contributions of each team member from time to time and evaluate the group's performance to continue the actions that contribute to the group's goal achievement and change inefficient measures (Johnson, 2003; Johnson & Johnson, 2009). In other words, group processing clarifies how the group is moving towards achieving their shared goal. In light of social interdependence theory, peer assessment can assist the team members to reflect on their performance and improve, particularly if implemented continuously throughout a team work project (Tucker et al., 2009). Including peer assessment in the process of the evaluating teamwork has also shown to be effective in enhancing students accountability and inspiring them to improve their interaction with the team members to achieve the team's goal (Scofield, 2005). It also highlights the individual contributions of the team members', thus motivating them for better performance.

In this work, we implemented a peer assessment process to individualise group marks, in engineering design courses, namely Professional Engineering Project Part A & B (Post Graduate course) and Engineering Design Project Part A & B (Under Graduate course), offered in the School of Engineering, Royal Melbourne Institute of Technology (RMIT), Melbourne, Australia. The peer assessment process was introduced to the teamwork assessment process in response to the course evaluation feedback that we received from the students in 2017. At the time, all team members received a single mark for the teamwork-based assessment tasks undertaken in these courses. However, their qualitative feedback suggested that they thought their teamwork marks did not account for "free-riders" thus not reflecting their true individual contributions to the completion of the task. The analysis of qualitative and quantitative student feedback both before and after the implementation of the peer assessment process demonstrated that it provides a fair approach to individualize the group marks and improves students' course experience.

## Materials and methods
### Course overview

Professional Engineering Project Part A & B and Engineering Design Project Part A & B courses are offered in undergraduate and postgraduate programs in Electrical Engineering, Electrical and Electronics Engineering, Computer and Network Engineering, and Communications Engineering. These courses run in two semesters, and their purpose is to introduce the students to engineering projects and refine their analytical and practical design skills through teamwork. The students engage in a team project (each team with 4-6 member) under supervision of an academic staff, leading to the design and production of an engineering product. In semester one, the students work in a set team to develop an Electrical, Electronic, Network, or Communication Engineering product idea. The teams are expected to complete their project definition, requirement analysis, and make significant progress in terms of designing and building their product. Teams complete their project in semester two, where they demonstrate their working prototype/product at the School of Engineering Trade Fair, which is usually attended and judged by many industry representatives.

**Assessment marking**

There are four tasks to be completed in Part A and three tasks in Part B. In Part A, teams are assessed for project definition, requirement analysis, presentation and progress report. The assessments in Part B include completion plan, presentation and final report. Marking of each assessment involves three steps. First, academic staff mark the assessment tasks, and each team receives a single mark for each task. Team members also assess their peers' performance after each assessment task. The marks received from the academics and the results of the peer assessments are then used to individualise and finalise the marks of each team member.

**Peer assessment and mark individualisation procedure**

Marks of individual students is determined by taking into account their individual peer assessment marks provided by their team members. This is achieved by inviting all team members to participate in the peer assessment process. However, participation in the peer-assessment process is not compulsory (See Appendix 1 for the peer assessment criteria). After the completion of each assessment task, students are given a week to peer-assess all team members including themselves in terms of their contribution to the completion of the assignment. Peer assessment is conducted using an on-line tool (SparkPlus ("https://sparkplus.com.au/," 2020) supported by RMIT University) as online peer assessment systems are reported to be more efficient (Tucker et al., 2009). The average of the peer assessment marks received by a student from everyone in the team is then calculated. The mark the teams received from the academics is then individualized proportionally based on the average of the peer assessment mark each student received from the team members. If a student receives zero average peer assessment mark, their individual mark for that assignment will be zero regardless of the mark given to the team by the academic staff. Indeed, the zero mark from all team members indicates that the student had not any individual contribution to the completion of the task. Table 1 illustrates an example of how the team marks are individualised based on peer assessments.

**Table 1:** The table shows an example of how a group assessment mark is individualised based on peer assessment marks. The team received 75 (out of 100) from the academic for their assignment. Only four team members out of 5 participated in the peer-assessment and Student 2 did not complete the peer assessment. The team's mark is individualised proportional to the average of peer assessment marks given to each student. In this example, individual peer assessment marks ranges from 70% to 90%. Accordingly, individual marks vary from 80.36/100 to 62.5/100. The individual marks are proportional to the peer assessments, for example, individual mark of student 1 is 80.36, which is obtained by multiplying 75 (team mark received from the academic) by 90 and then dividing by 84 (the average of all team members' peer assessment marks).

|  | Peer assessment completed for: | | | | |
|---|---|---|---|---|---|
| *Name of Assessor* | **Student 1** | **Student 2** | **Student 3** | **Student 4** | **Student 5** |
| *Student 1* | 100% | 60% | 100% | 80% | 40% |
| *Student 2* | - | - | - | - | - |
| *Student 3* | 100% | 100% | 100% | 100% | 100% |
| *Student 4* | 80% | 80% | 80% | 100% | 60% |
| *Student 5* | 80% | 80% | 80% | 80% | 80% |
| ***Mean peer assessment mark*** | ***90%*** | ***80%*** | ***90%*** | ***90%*** | ***70%*** |
| ***Team mark from the academic*** | ***75*** | ***75*** | ***75*** | ***75*** | ***75*** |
| **Individual mark** | 80.36 | 71.43 | 80.36 | 80.36 | 62.5 |

## Course evaluation

Students' written feedback collected through the Course Evaluation Survey (CES) (see Appendix B for CES procedure) was used to investigate the impact of implementing peer assessment process to individualise the team member marks on the students' course evaluation. We analysed student feedback coming from two groups of students. The group who provided feedback before the implementation of the peer assessment process, included 151 students (27 teams) enrolled in the Professional Engineering Project and Engineering Design Project courses in 2018. Then, we implemented the peer assessments process in 2019 and the feedback collected from 129 students (23 teams) enrolled in 2019 was also analysed to investigate how the changes to the assessment process impacted the students course evaluation.

# Results

In this section, we present our quantitative and qualitative results pre and post implementation of the peer assessment strategy for the fair individualization of team marks.

## Pre peer assessment

In this case, the students' individual contributions to the teamwork were not assessed, and a single mark was given to each team. In the written feedback provided in the Course Evaluation Survey (CES) report, 24% of the students provided feedback in the "best aspects of the course" section appreciated teamwork aspects of the course (Table 3). When students were asked for the aspects of the course that needed improvement, 26% of them explicitly mentioned the marking process and how individual contributions were captured as an area for improvement (Table 3). They particularly mentioned social loafing as a major issue and suggested that to be taken into account in the assessment process. The following quotations are examples of the student feedback in this regard:

> *"There's no consequence for not contributing to the group assessments";*
> *"It makes it way too easy for group members to be lazy and let the rest of the group do everything";*

> *"As with any group work, two students allocated to our group completed little to zero work all year but will receive the same mark as the rest of us";*
>
> *"Team mates not pulling their weight and not being marked down enough for it".*

To respond to the students' feedback, we implemented the peer assessment process to individualise the team marks.

**Post peer assessment**

The students were asked to complete peer assessments after each assignment. Analysis of the qualitative feedback provided in the CES survey (see Appendix B) showed that there was a 42% increase in students' positive feedback indicating their appreciation of the teamwork aspect of the assessment tasks compared to their feedback received prior to the implementation of the mark individualisation process (Table 3).

**Table 3:** Impact of group mark individualisation based on peer assessment policy in the feedback provided by students in the CES survey.

| Pre peer assessment ||
|---|---|
| % of student commented about teamwork appreciation in the section "What are the best aspects of this course?" of the CES | % of students commented on inappropriate team assessment in the section "What aspects of this course are in most need of improvement?" of the CES |
| 24% | 26% |
| Post peer assessment ||
| 34% | 7% |

Almost one-third of the students who provided CES feedback, explicitly mentioned that they liked the teamwork aspects of the course. The following quotations are instances of student feedback reflecting their appreciation of teamwork skill development in the courses after the implementation of the peer assessment process to individualise their marks:

> *"(I liked) being in groups which helps us in our communication and also providing feedback and tips to do the given module is pretty good.";*
>
> *"Learn about the process of how to start a project, and team cooperation."*
>
> *"Learn how to assign tasks and how to play the role of the team to complete the task."*

*"Working in a project with different people gives training on how to perform in a group. Mainly the management of group and working together in a group. It was a good experience."*
*"I learned working in team and problem solving skills.";*

*"This course helps in building the Industrial Network and team building".*

As reflected in these comments, the students appreciated the improved group processing as a result of implementing the peer assessment process. This highlights the importance of group processing (Johnson & Johnson, 2009) in improving the cooperation of team members.

Teaching quality of the courses at RMIT are assessed through Good Teaching Scale (GTS) in the CES survey (see Appendix B). Teams were asked to complete the CES for their academic supervisors. Our data showed that the post peer assessment mean GTSs (mGTSs) were significantly improved ($P < 0.018$; Wilconxon's ranksum test) compared to the pre peer assessment mGTSs (Table 4). The peer assessment process improvement the mGTS scores by 8%, while other courses offered in the school and RMIT experienced less than 1% improvement over the same time period. The Overall Satisfaction Index (OSI) of the course was also increased from 4 to 4.12 by implementing the peer assessment. The improvement in the mGTS and OSI provide evidence for the positive impact of the peer assessment process and individualising the team scores.

**Table 4:** Impact of group mark individualisation on the student engagement with the course measured by mGTS and OSI (see appendix B for details). mGTS shows the average and standard deviation of the mGTSs of the academic staff involved in supervising the student projects. mGTS of post peer assessment is significantly higher than that of pre peer assessment ($P < 0.018$; Wilconxon's ranksum test).

| mGTS | OSI |
|---|---|
| **Pre peer assessment** | |
| 4.04±0.47 | 4 |
| **Post peer assessment** | |
| 4.34±0.46 | 4.12 |

## Discussion

Efficient teamwork skill is essential for many engineering jobs. Many accreditation agencies, like Engineers Australia that is the peak accreditation body for engineering degrees in Australian universities, expect engineering graduates to gain proficiency in teamwork skills in their education programs. To be truly proficient in teamwork skills, students must be offered courses with teamwork as their core. However, assessing teamwork in challenging. A critical challenge is how to effectively individualise team marks. Engineering team projects often result in a single end-product, receiving a single mark from assessors.

In this work, we implemented a peer assessment process as a part of assessing engineering design courses offered in undergraduate and postgraduate programs of Electrical and Electronics Engineering at RMIT University, Melbourne, Australia. This was done to individualise team assessment marks in response to the student feedback about the shortcoming of all team members receiving the same mark for the team project. After completing each of the 7 parts of the assignment, students were asked to complete a peer assessment on an online platform (SparkPlus) to evaluate their team members' performance. The average rating that each student received from their team members were then used to individualise the team mark for that student. This resulted in fair individual distribution of marks accounting for the individual contributions in the marking process. Our findings showed that the students' evaluation of the course significantly improved after the implementation of the peer assessment process and considerably decreased student complains about marking fairness for non-contributing team members. The peer assessment was also a major contributor to the improved good teaching scores for the academic supervisors involved in supervising the team projects. The data also showed that students' engagement with the course was significantly improved. The analysis of the students qualitative feedback reflected that the peer assessment process improved the group processing as a factor mediating effective cooperation (Johnson & Johnson, 2009) which resulted in student's having a greater appreciation of the teamwork aspect of the assessment.

Our findings confirms previous research in showing that peer assessment is an effective way of individualising team marks and accounting for the individual contributions of the team members and preventing team members from taking advantage of freeriding (Gransberg, 2010; Tucker et al., 2009). Students benefit from the team-based assessment by improving and enhancing their critical employability skills. At the same time the peer assessment and individualising their marks highlights their individual contributions to the accomplishment of the project and encourages them to avoid freeriding.

# Appendices

## Appendix A: Questions asked in the peer assessment enquiry

Each student is assessed by their team members for contributions to the team activity, overall performance and level of involvement with the project works. The team members are assessed against the following criteria:

- Ethical conduct and professional accountability. **(20%)**
- Creative, innovative and proactive demeanour. **(20%)**
- Efficiency use and manage information. **(20%)**
- Orderly manage her/himself. **(20%)**
- Effective team member. **(20%)**

Students provide a rating of 0-10 for each of the criteria above for each of their team members, which is then averaged over all criteria, and multiplied by 10 to provide the percentage (0%-100%) for each peer assessment. When logged in to the online system (SparkPlus), students find the assignment to which the peer assessment is through. Then for each criterion above, they find peers to whom they require to assess. Students can also provide written feedback to each of their peers in the team. The identity of peer assessors is blind to the peers. As the peer assessment period concludes, the average peer assessment received by each student along with the written feedback provided by peers is published and becomes accessible to students in their portal.

## Appendix B: Course Evaluation Survey Questionnaire

In the last four weeks of each semester, students are asked to complete a Course Evaluation Survey (CES) for each course enrolled. They provide written feedback on their experience of the course and also provide a rating of 1-5 for each of the following questions:

Q1. The teaching staff are extremely good at explaining things.
Q2. The teaching staff normally give me helpful feedback on how I am going in this course.

Q3. The teaching staff in this course motivate me to do my best work.

Q4. The teaching staff work hard to make this course interesting.

Q5. The staff make a real effort to understand difficulties I might be having with my work.

Q6. The staff put a lot of time into commenting on my work.

Q7. Overall, I am satisfied with the quality of this course.

The ratings for each CES item above are averaged over all students provided valid answers. Then, the averaged ratings of Q1-Q6 are again averaged to obtain mean Good Teaching Scale (mGTS) that is used for different policy-making actions, such as academic staff promotion and external block funding approvals. The average rating of Q7 is used to obtain Overall Satisfaction Index (OSI). The course receives a single mOSI, while each academic supervisor involved in the course receives an mGTS.

The students are also asked to provide feedback on positive and negative aspects of the course. Students provided written feedback in CES surveys on two questions: "What are the best aspects of this course?" (positive aspects) and "What aspects of this course are in most need of improvement?" (negative aspects). The written feedback is used to make positive changes in the course to improve course quality and student experience.